\documentstyle[11pt,newpasp,twoside,psfig]{article}
\reversemarginpar

\begin{document}

\title{X-ray Transients \hfil\eject
from X-ray Binaries to Gamma Ray Bursts}
\author{John Heise$^{1,2}$ and Jean in 't Zand$^{2,1}$}
\affil{$^1$Space Research Organization Netherlands, Utrecht, NL-3484CA,
$^2$Astronomical Institute, Utrecht University, Utrecht, NL-3508 TA,
Netherlands}

\begin{abstract}
We discuss three classes of x-ray transients to highlight three new
types of transients found with the Wide Field Cameras onboard BeppoSAX. 
First there are the transients related to Low Mass X-ray Binaries 
in outburst, typically lasting weeks to months and reaching luminosities
of the Eddington limit for a few solar masses. Recently another 
subclass of outbursts in such binaries has been discovered, which
are an order of magnitude fainter and last shorter than typical 
hours to days.
We discuss whether they constitute a separate subset of x-ray binaries.

A second class of x-ray transients are the x-ray bursts.
Thermonuclear explosions on a neutron star (type~I x-ray bursts)
usually last of order minutes or less. 
We discovered a second type (called super x-ray bursts) with a duration of 
several hours. They relate to thermonuclear detonations much deeper in 
the neutron star atmosphere, possibly burning on the nuclear ashes of 
normal x-ray bursts.

The third class are the enigmatic Fast X-ray Transients occurring
at all galactic latitudes. We found that the bright ones are of two types only:
either nearby coronal sources (lasting hours) or the
socalled x-ray flashes (lasting minutes). The new class,
the X-ray flashes, may be a new type of cosmic explosion, intermediate between
supernovae and gamma ray bursts, or they may be highly redshifted gamma ray bursts.

It thus appears that the three classes of x-ray transients each 
come in two flavors: long and short.
\end{abstract}

\section{Introduction}


\begin{figure}[h]
\centering
\hbox{
\psfig{file=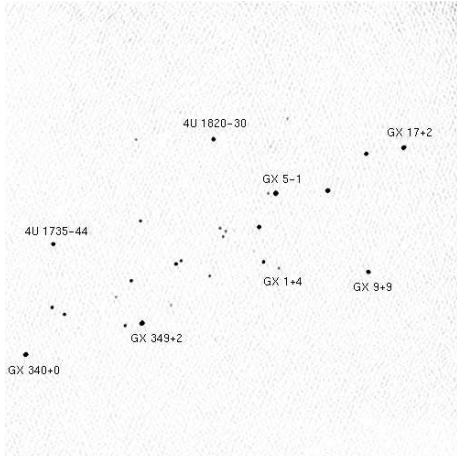,width=0.45\textwidth,clip=t}
\parbox[b]{.5\textwidth}{\caption[]
{A ty\-pi\-cal WFC ima\-ge of $40^o\times 40^o$  a\-round the
ga\-lac\-tic cen\-ter. 
The ang\-ul\-ar re\-so\-lu\-tion is 5\arcmin.
This field con\-tains about $50\%$  of all LMXBs in the gal\-axy.
Po\-si\-tion de\-ter\-mi\-na\-tions have an ac\-cu\-ra\-cy of 
ty\-pi\-cal\-ly 1\arcmin -3\arcmin. 
About 30 point sour\-ces are seen ab\-ove the de\-tec\-tion thres\-hold
of ty\-pi\-cal\-ly 10 mCrab in $10^5$ s.
\vspace*{3mm}
}
}
}
\label{wfcGC}
\end{figure}

The x-ray sky is transient in nature. X-ray sources appear and disappear.
This fact has been known since the beginning of x-ray astronomy,
soon after the discovery of the first few x-ray sources. 
Before space born instruments, x-ray detectors
were placed on sounding rockets and scanned the entire sky in a spinning
payload during a few minutes above the atmosphere. 
On the time scale of successive launches
(months to years) bright x-ray sources may vanish below detection limits and
new sources may be among the brightest x-ray sources.  
Later, variablity on all time scales (millisec to years) became a well known 
characteristic of the x-ray sky. 
For x-ray binaries, many of these time scales are now explained
in terms of accretion onto compact objects. 
In this review I limit the discussion on these transients 
to two, recently discovered, phenomena: 
faint transients and x-ray superbursts.

An entirely different class of transient phenomena are the 
Fast X-ray Transients (FXTs) or High Latitude Transients.
They were discovered (Forman et al. 1978, Pye \& McHardy 1983,
Ambruster et al. 1986) with the first few satellite born x-ray instruments, 
such as those on board UHURU, Vela, Ariel V, HEAO-1, etc. 
These rotating satellites scanned the entire sky.
The scan frequency often occurred on the same time scale as the satellite orbit,
typically 1.5 hours. 
Such events are detected in one sky scan and disappeared soon afterward,
typically limiting the duration to be longer then a minute and shorter than
a few hours. 
In some cases the FXT was detected on the next sky scan as well,
setting a duration of order one hour.
Because of the limited amount of data, the nature of these sources,
although seen by almost every x-ray satellite, remains an enigma. 

The attitude mode and orbit of the satellite continues to determine
the character of the time scale of x-ray sources.
3-axis stabilized x-ray satellites made the discovery possible of
short and sparsely distributed outbursts in known sources.
The source monitoring is only interrupted by earth occultations. 
The first of such a satellite was the Astronomical Netherlands Satellite ANS.
With the x-ray instruments on ANS the first x-ray burst was
discovered in 2U1820-30 (Grindlay \& Heise 1975, Grindlay et al. 1976).
They later later appeared to be thermonuclear explosions in the
accreting layers of neutron stars in x-ray binaries.
We discuss the recently discovered new type of thermonuclear
explocions in section 3.

\section{Bright and dim X-ray transients from X-ray Binaries}
\begin{figure}
\hbox{
\psfig{file=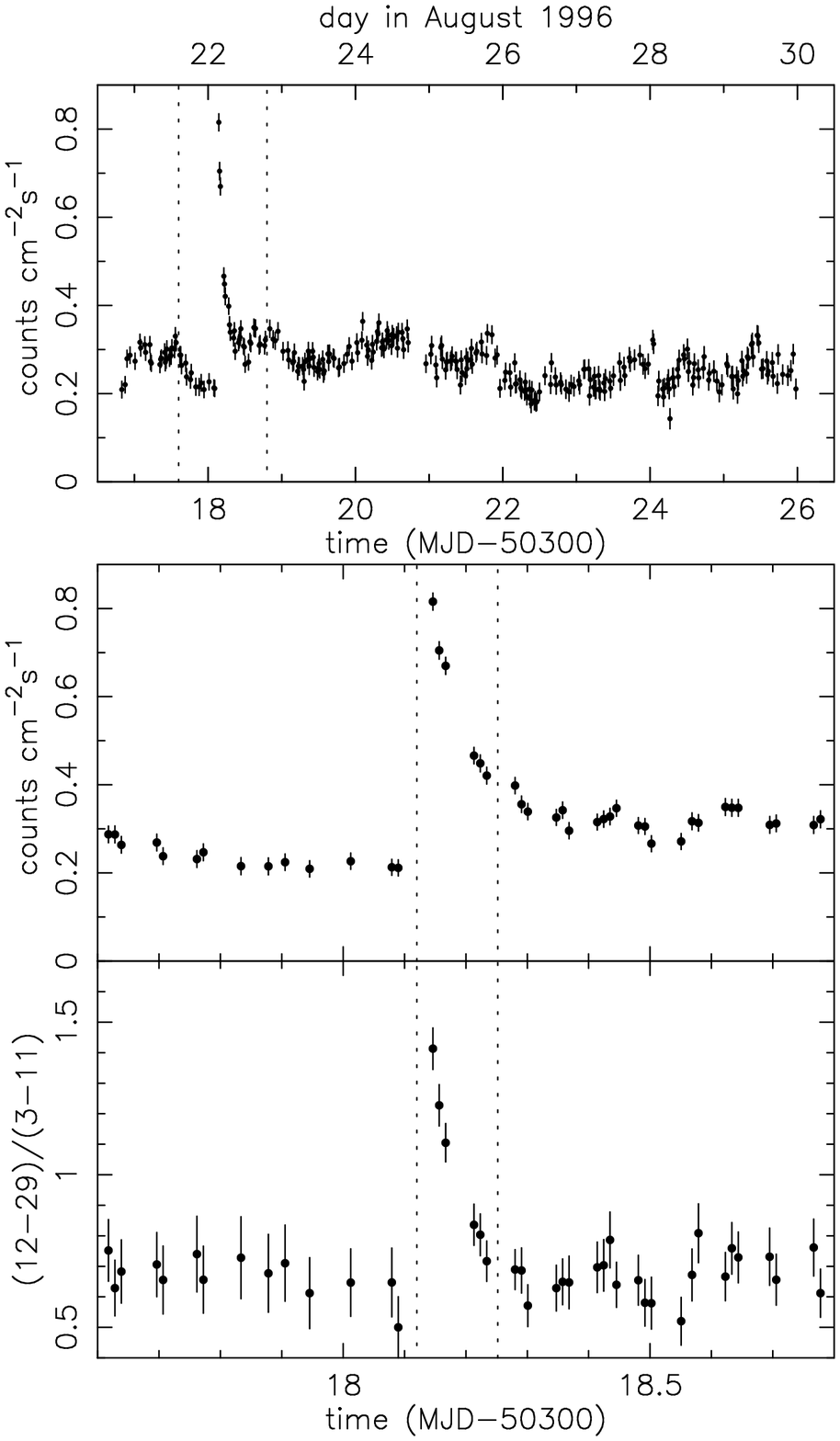,width=0.4\textwidth}
\hspace*{0.025\textwidth}
\psfig{file=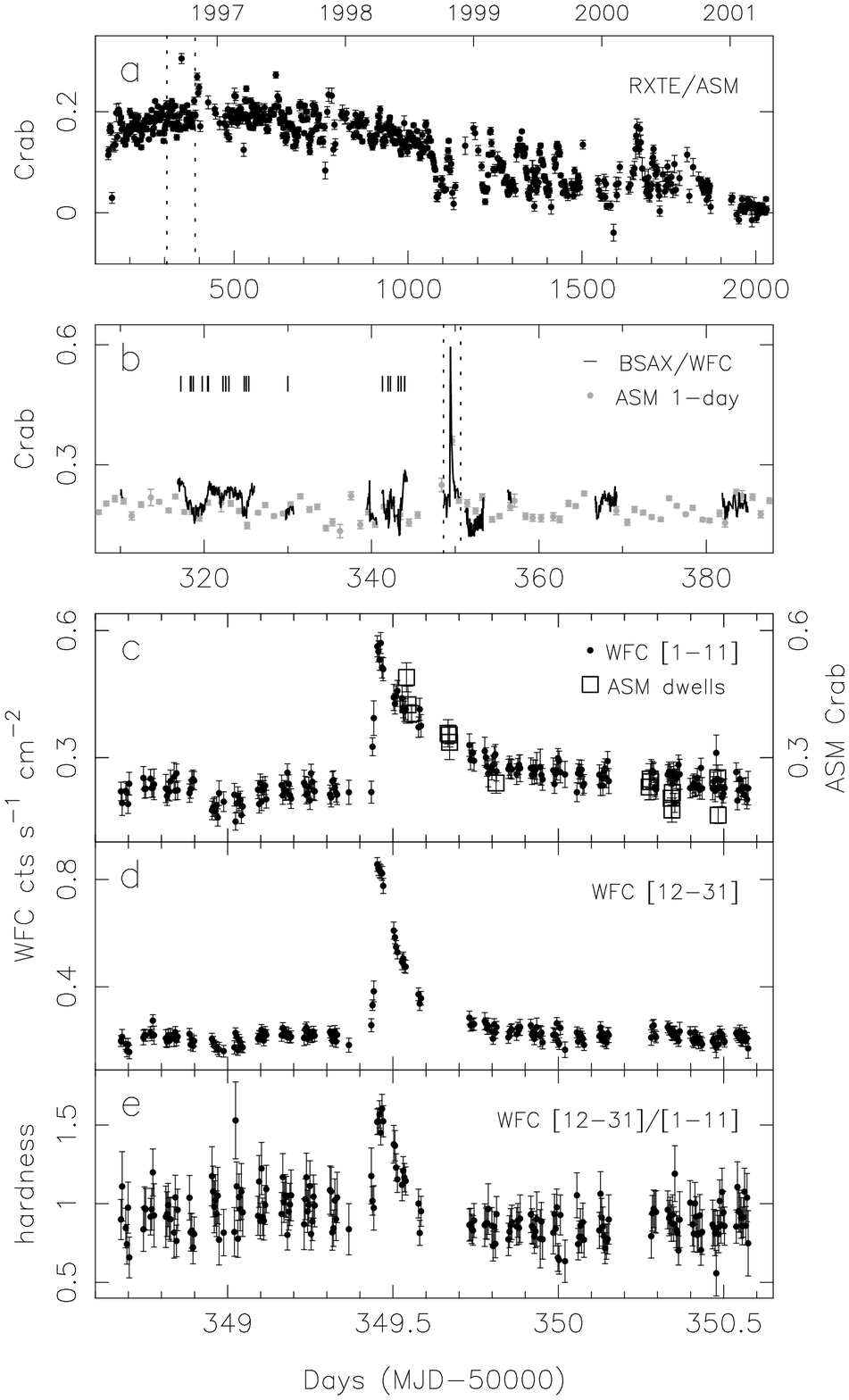,width=0.4\textwidth}
}
\caption
{{\it Left:} Discovery observation of a superburst in 4U1735-44.
From Cornelisse et al. (2000).
{\it Right:} The superburst in KS1731-260 lasting for 12 hours.
From Kuulkers et al. (2001).
Both are discussed in the text. 
}
\label{superbursts}
\end{figure}

Two categories of x-ray binaries are high-mass x-ray binaries
(HMXBs) and low-mass x-ray binaries (LMXBs).
HMXBs are young systems in which a star of spectral type O or B has a
compact companion star. 
They are concentrated toward to galactic plane.
In LMXBs binaries the companion star has spectral type
A or later. 
They are older and the spatial distribution is concentrated
toward the galactic center. 
A large fraction of them are bright transient
x-ray sources (see e.g. van Paradijs 1995). 
More than half of the LMXBs  is found within 20$^o$ of the galactic center. 
They have been studied by all sky x-ray monitors 
(such as the RXTE ASM, Bradt et al. 1999),
by wide field coded mask  x-ray instruments 
(such as ART-P and SIGMA on the Granat platform and COMIS on the KVANT module). 
Two currently active programs use the Proportional Counter Array (PCA) on 
RXTE in a scanning mode and the $40^o \times 40^o$ full width zero 
response field of the BeppoSAX Wide Field Cameras (WFCs).
In the WFCs all sources are simultaneously monitored, see e.g.
Fig.~1.  
In recent years the number of new transient LMXBs has been expanded by $25\%$.
A review of recent x-ray transients in the galactic center region
is given by in 't Zand (2000).
The BeppoSAX and RXTE observations show that there is are a significant
number of transients that do not reach high peak luminosities:
many of the newly found LMXB transients are faint. 

\begin{figure}
\hbox{
\psfig{file=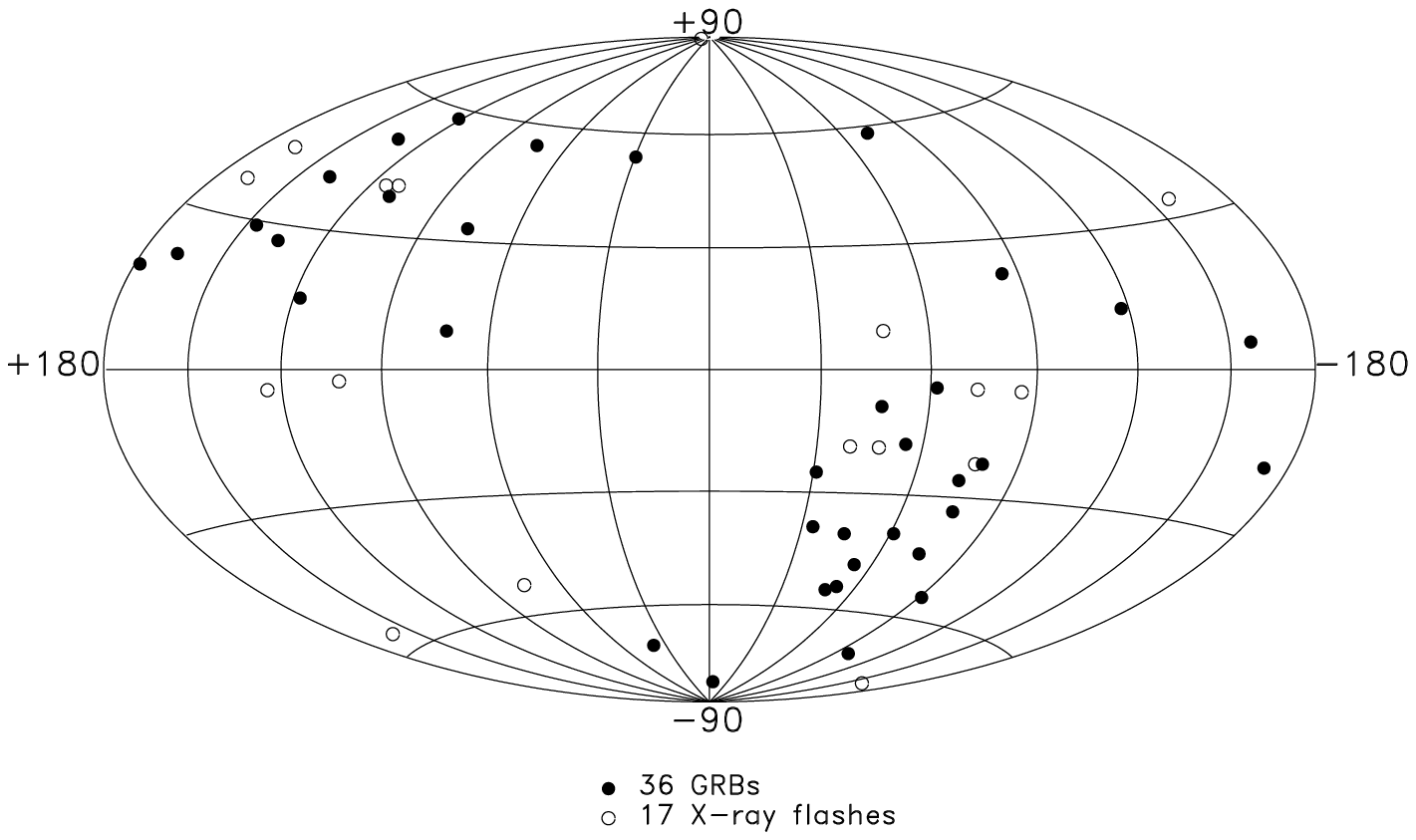,width=0.45\textwidth}
\psfig{file=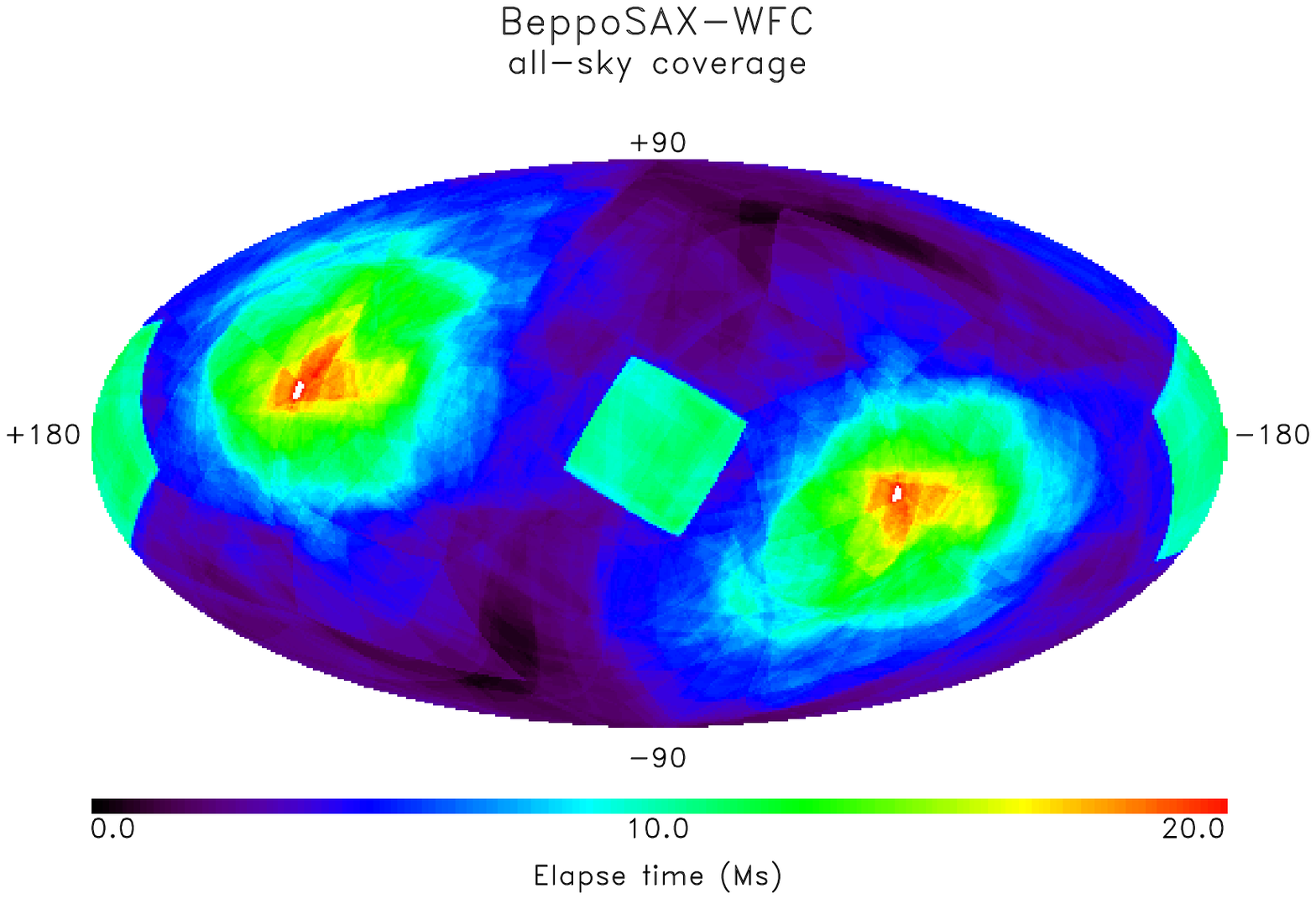,width=0.45\textwidth}
}
\caption
{{\it left:} Spatial distribution of WFC-detected x-ray flashes (open circles) 
and x-ray counterpart of GRBs (filled circles) in galactic coordinates. 
The distribution is consistent with isotropy for the given sky coverage.
{\it Right:} Total sky coverage of WFC observations with BeppoSAX. It shows
that the observed sky is biased towards the Galactic Center region and 
two regions perpendicular to it (since the two WFCs are mounted perpendicular to
the Narrow Field Instruments).
}
\label{spatial}
\end{figure}

The large simultanous exposure of many sources in the galactic center region
with the WFC increases the chance of observing type~I x-ray bursts. 
Indeed, many of the newly found transients (Heise et al. 1999) exhibit 
such thermonuclear bursts,
identifying the compact star in the binary system as a neutron star
and excluding a black hole hypothesis. 

One of the faint transients discovered in
the WFC is the binary accreting x-ray pulsar SAX J1808.4-3658 
(in 't Zand 1998, 2000a).
Heise et al. (1999)  propose that the faint transients
make up an as yet unrecognized subclass of LMXBs, characterized by
a neutron star as a compact star and a special nature of the companion star.
For instance, the companion star in SAX J1808.4-3658 has a very low mass
(Chakrabarty \& Morgan 1998). 
The existence of a population of faint transients in the galactic center
and dominated by neutron stars as compact object
would be consistent  (King 2000) with the model of an irradiated
accretion disc as the cause of the instability causing the transient
behaviour.
Unfortunately, the evidence that the subclass of faint transients 
relates to a special kind of x-ray binary, is not mounting.  
In the galactic center region, unraveling the binary orbit and 
identifying the optical counterparts of these transients is difficult.

\section{Normal and super thermonuclear outbursts on Neutron Stars}

Normal thermonuclear (so called type~I) x-ray bursts are reviewed by
Lewin et al. (1995). The x-ray light curve is
characterized by a fast rise followed 
by a decay on a time scale from seconds to minutes.
The x-ray spectra are thermal and can be fitted with black bodies.
The characteristic temperature decreases in the later 
stages of the burst. This softening is interpreted as cooling after 
the onset of a sudden thermonuclear instability.
The explosion is though to arise in the sediments
of the accretion flow in the atmosphere of an accreting neutron star.

\begin{figure}
\centering
\hbox{
\psfig{file=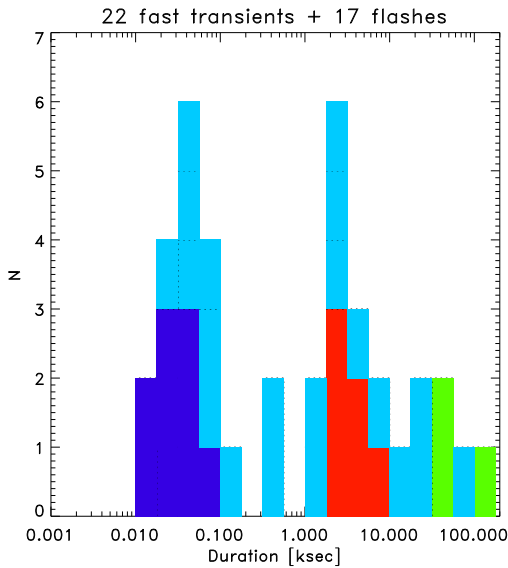,width=0.50\textwidth}
\parbox[b]{.50\textwidth}{\caption[]
{Histogram of durations of FXTs seen with the WFC, 
showing two populations: 
one with a duration of 0.5 to 20 min,
and one lasting typically an hour and longer. 
The long duration class contains identifications with coronal systems 
(flare stars marked dark grey and RS CVNs marked
lighter grey), the short duration class (0.5 to 20 min) are called
x-ray flashes (XRFs. The flashes for which there exist a $20-100$ keV flux
9seen with BATSE) are marked black.\vspace*{3mm}
}
}
}
\label{FXTs}
\end{figure}

A new kind of x-ray outburst has been seen in the same sources
that exhibit normal thermonuclear bursts. 
It concerns exceedingly long bursts, which are called super x-ray bursts. 
These bursts are extremely rare, and it is mainly thanks to the {\em long} and
{\em population-wide} WFC coverage of LMXBs
that they are picked up. 
The first one was discovered (Cornelisse et al. 2000) 
with WFC in 4U~1735-44 in 1996 data. 
Fig.~2 (top left) shows a 9 day light curve with the WFCs (2-28 keV in 15 min bins).
The middle and lower panel shows an enlarged version together with
the hardness ratio (5-20 keV/2-5 keV), showing the typical softening
of the burst as one of the main characteristics of
thermonuclear x-ray bursts. Except for the duration and the total fluence,
the superburst is similar to type~I bursts.
The fluence and duration is 300 times larger than the longest burst 
previously seen in this source.

Two more examples of superbursts have been found in WFC data,
in Ser X-1 with a duration of 4 hours (Cornelisse et al., 2001) and 
in KS 1731-260 (Kuulkers et al., 2001). 
The superburst in KS1731-260 is seen in Fig.2. The two top panels
show a long duration light curve in both ASM and WFC data. 
Times of normal type~I x-ray bursts are indicated with vertical lines
in panel b, and only seen {\it before} 
the superburst. The superburst lasts for 12 hours, the longest so far.
Panel c and d show the light curve in 2-5 keV and
5-28 keV respectively, with the hardness ratio in panel e.

The phenomenon is also in RXTE observations. 
A very well defined example has been found in PCA data
from 4U~1820-30, a binary with a white dwarf companion 
(Strohmayer et al. 2001), and 4U~1636-53 (Wijnands 2001).
The recurrence of these
hours-long bursts has been detected in ASM data from 4U~1636-53 (Wijnands
2001). The long duration, two to three orders of magnitude
larger than 'ordinary' bursts, clearly challenges the theory of thermonuclear
flashes. Two possibilities exist. Either unstable carbon burning using
the remaining ashes after normal thermonuclear bursts, or another process is 
responsible for the energy release. Different mechanisms may be responsible
for different super bursts. 
Kuulkers et al. (2001) propose for  the superburst from KS1731-260
a runaway electron capture, via the $p(e^-,\nu)n$ reaction,
occurring at densities where the Fermi energy
of the electron becomes larger than the mass difference between
neutrons and protons.

\begin{figure}
\begin{center}
\hbox{
\psfig{file=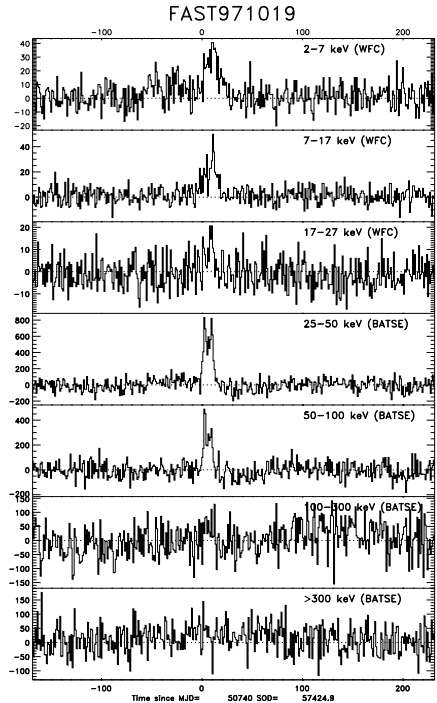,width=.5\textwidth}
\hspace*{.05\textwidth} 
\parbox[b]{.45\textwidth}{\caption[]
{XRF 971019, 
the X-ray flash with the larg\-est X-ray flu\-ence.
The light curve is shown in three dif\-ferent ener\-gy bands of the Wide
Field camera and also in the low\-est chan\-nels of BATSE 1 sec data. 
It was not a trig\-ger\-ed GRB. Most x-ray flashes only show up in
the lowest BATSE channel.}\vspace*{4mm}}
}
\label{flashdata}\end{center}
\end{figure}

\section{Fast X-ray Transients/High-latitude X-ray Transients}

Fast X-ray Transients (FXTs) have been observed with many
x-ray satellites. Usually they are loosely defined as
x-ray transient events that occur on a time scale of 
less than a few hours and that are not related to x-ray binary systems.  
FXTs thus contrast in duration with the previously discussed 
transient x-ray sources in LMXBs which typically have 
durations longer than a day.
FXTs occur at high galactic latitudes and therefor are also
sometimes called High Latitude X-ray Transients.
The data on these serendipitous sources is often
limited. 
As a consequence not much is known about the origin of FXTs.
Indeed, there is not even an a priori reason for a common origin.
FXTs may be a mixture of several types of astronomical events 
and (in some cases) x-ray detector artifacts.
Observations with the WFCs in its 5 years operational period
have a sufficient sky coverage and allow for the
determination of duration, lightcurves and spectra,
while the positional information is sufficiently good to search
for identifications.  

\subsection{Historical overview of Fast X-ray Transients}

The first FXT of this type was seen with
UHURU (Forman et al. 1978). 
We review some of the other observations.

\subsubsection{The Ariel V sky survey of Fast Transient X-ray sources.}  

Ariel V scanned the sky for 5.5 years with a time resolution of one 
satellite orbit ($\sim100$ min) and Pye \& McHardy (1983) reported
27 events.  
The distrubution is consistent with isotropy.
About 20\% of the sources seen
with Ariel V are identified with active coronae in RS CVn systems 
and have a duration of order hours. 
The authors conclude that all Ariel V observations are consistent 
with as yet unknown coronal sources, but remark that two of the transients 
are time coincident with gamma ray burst sources. 
One of them also was spatially coincident with a GRB to within
$\sim 1^o$. 
The frequency is estimated as one FXT every $\sim 3$ days above 
$4\times 10^{-10}$ erg/s/cm$^2$ (2-10 keV).

\subsubsection{The HEAO-1 all-sky survey of fast X-ray Transients.} 
HEAO-1 also had complete sky coverage. 
Ambruster et al. (1986) report the analysis of the first 6 month of
HEAO-1 scanning data. They observe 10 FXTs with the A1 instrument
(0.5-20 keV) above $\sim 7\times 10^{-11}$ erg/s/cm$^2$, 
of which 4 are identified with coronal systems:
3  flare stars and 1  RS CVn systen.  
The duration is $> 10$ s and $< 1.5$ hr. 
On the basis of variability within
the observation of 10 s and the absence of apparent counter parts
in other wavelengths, Ambruster et al. suggest that one source 
could have been x-ray emission from a faint GRB.
They estimate an all sky rate of $\sim 1500$ per year. 
In the A2 instrument (2-60 keV), 5 more FXTs
have been detected (Connors et al. 1986)
with peak fluxes between $\sim 10^{-10}$ and  $\sim 10^{-9}$
erg s$^{-1}$cm$^{-2}$ (2-10 keV). 
Connors et al. suggest that most of the events are hard
coronal flares from dMe-dKe stars, with a flare rate of $2\times 10^4$ per year 
above $\sim 10^{-10}$ erg s$^{-1}$cm$^{-2}$ and 
they rule out the identification with GRBs.

\begin{figure}
\hbox{
\psfig{file=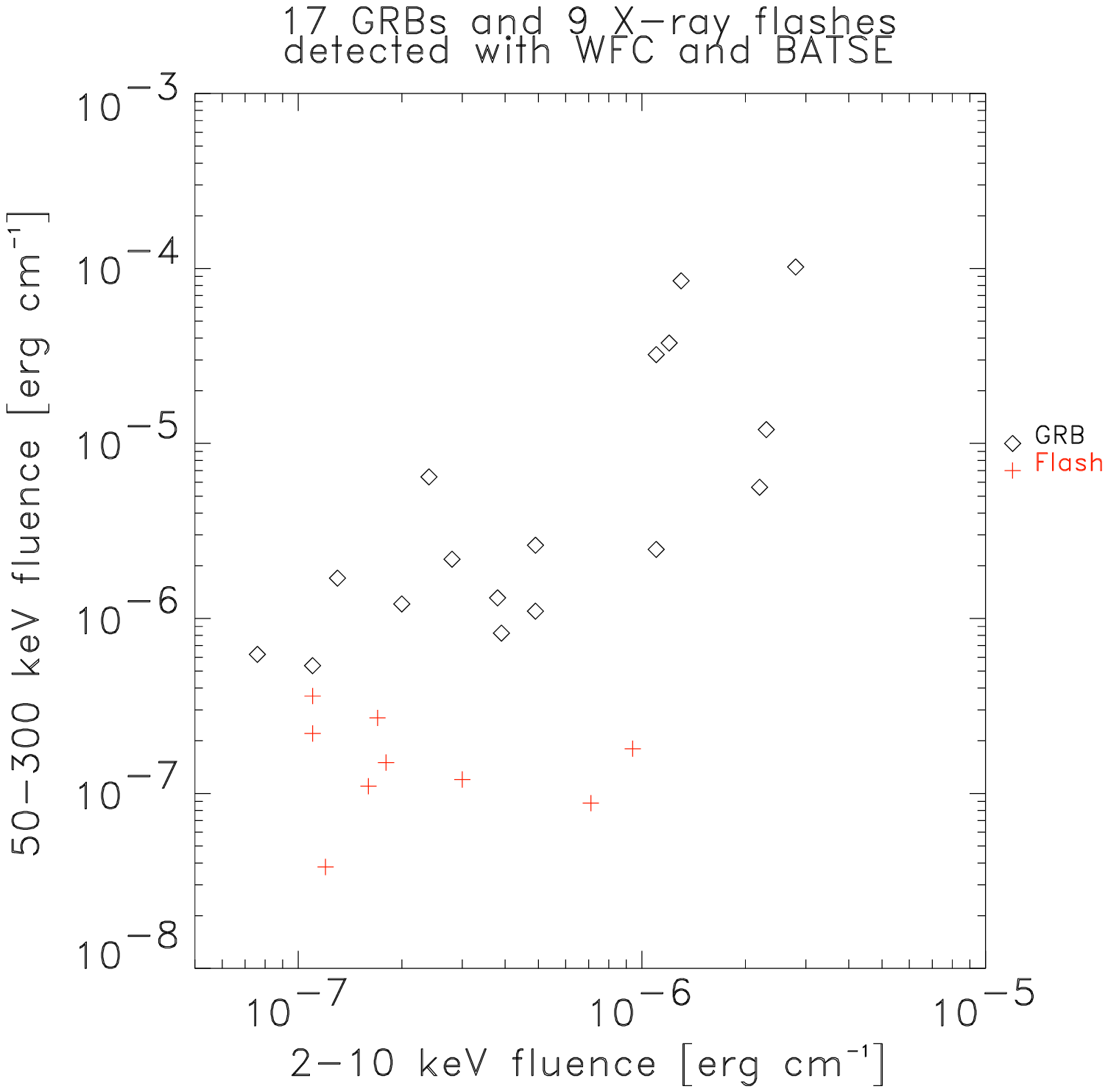,width=0.5\textwidth}
\psfig{file=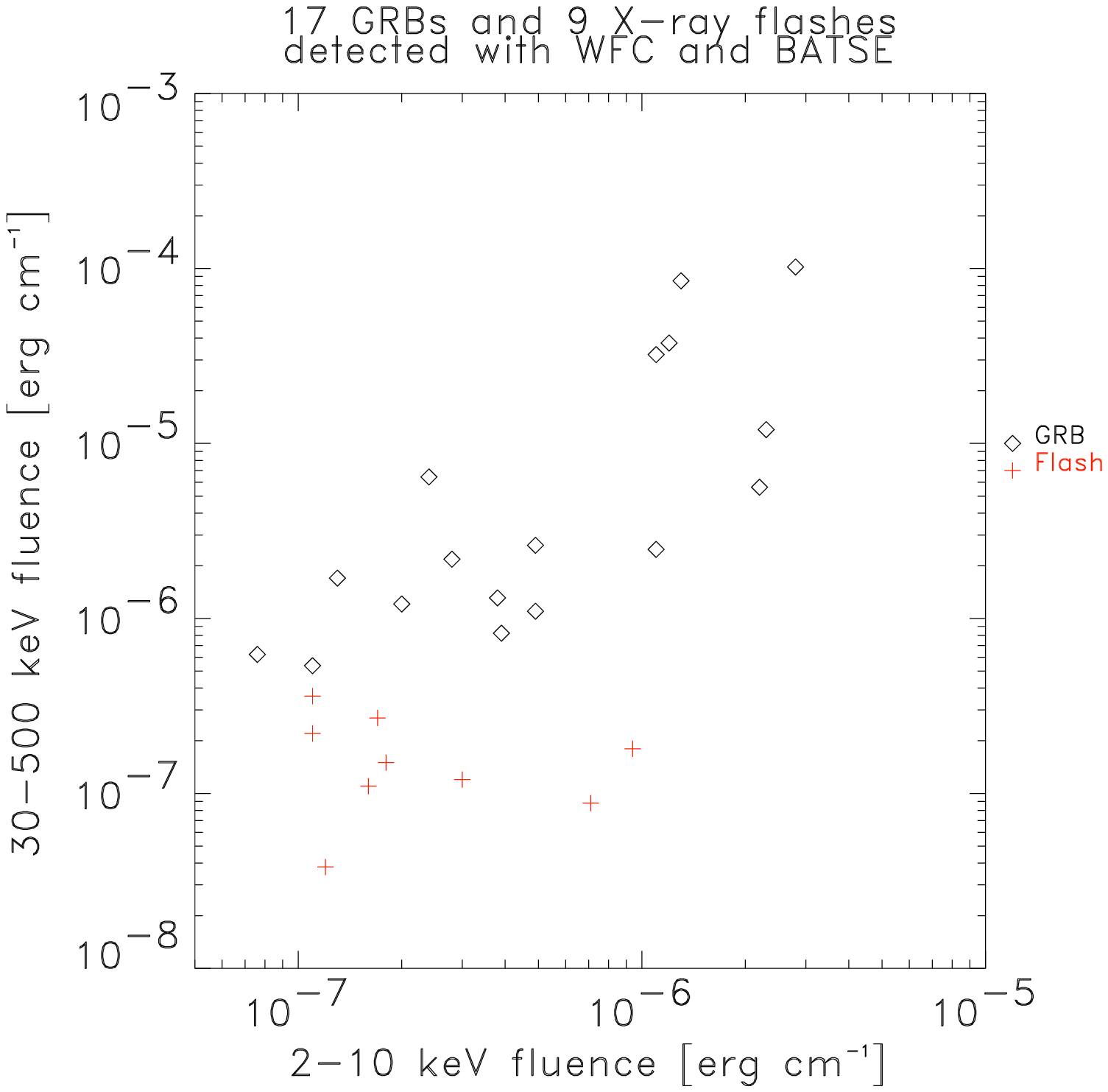,width=0.5\textwidth}
}
\caption
{The gamma ray peak flux versus x-ray peak flux ({\it left})
and the gamma ray fluence versus x-ray fluence ({\it right})
is shown for all short FXTs for which both quantities are measured
with x-ray fluxes from the WFC and $\gamma$-ray fluxes above 30 keV
from the 1 sec BATSE data.
They show the distiction between x-ray flashes (crosses) and GRBs (diamonds)
in the two ways. From Heise et al. 2000.
}
\label{gamma-x}
\end{figure}

\subsubsection{Einstein observatory detection of faint X-ray flashes.}

Gotthelf et al. (1999) searched for x-ray counterparts to GRBs in 
the data from the Imaging Proportional Counter (IPC) on-board the Einstein Observatory
to a limiting sensitivity of $10^{-11}$ erg s$^{-1}$cm$^{-2}$ in the
0.2-3.5 keV band. 
The totale exposure time is $1.5\times 10^7$ s.
On a time scale of up to $\sim 10$ s they find 42 events
of which 18 have spectra consistent with an extragalactic origin and light curves
similar to x-ray counterparts of GRBs, 
although many events are much shorter than 10 s.
The events are not identified in other wavelenghts on a one arc-minute spatial scale.
They are not correlated with nearby galaxy distribution.
The implied rate of $2\times10^6$ yr$^{-1}$ is
far more numerous than known GRBs. 
The flashes are found to be isotropically
distributed and have approximately Euclidean number-size relation.

\subsection{X-ray counterparts of GRBs.}
\subsubsection{X-ray spectral characteristics of Ginga gamma-ray bursts.} 

X-ray (1-8 keV) counterparts of GRBs were first detected in 
1973 (Wheaton et al. 1973). Stroh\-mayer et al. (1998)
summarize the results observed with the {\it Ginga} satellite in the range
2-400 keV. Out of 120 GRBs in the operational period of 4.5 years between
1987 March and 1991 October, 22 events were studied. 
The average flux ratio of the x-ray energy (2-10 keV) to the gamma ray energy 
(50-300 keV) is  0.24 with a wide
distribution from 0.01 to more than unity. Photon spectra are well described 
by a low-energy slope, a bend energy, and a high-energy slope.  The distribution
of the bend energy extends to below 10 keV, suggesting that GRBs might have
two break energies, one in the 50-500 keV range and the other near 5 keV.

\subsubsection{X-ray counterparts of GRBs in WFCs.} 

The Wide Field Cameras (WFCs) on board BeppoSAX (2-25 keV) 
combine a large field of view ($40^o\times 40^o$, 
full width to zero response) with a resolving power 
of 5 arcmin and allow for a fast position determination. 
Since the launch in 1996 about 50 X-ray counterparts of GRBs have been 
localized and studied. 
Typically error circles around the localization are between 2 and 3 arcmin
with 99\% confidence levels. These positions 
have been established after an average delay of 4 to 5 hours.
The error regions fall within the field of view of most optical, radio and
x-ray telescopes and have triggered the discovery of afterglows in these
wavelength bands. The average deviation between the WFC position and
the optical transients found is within the error circle radius and consistent
with statistics.

The average rate of  GRB prompt counterparts observations is
about 9 per year. The X-ray counterparts can be very bright and range
between $10^{-8}$ and $10^{-7}$ erg/s/cm$^2$. The average
spectra are characterized by a power law shape, with photon indices 
between 0.5 and 3 (see Heise et al. 2000, Fig.~2).  
The T90 durations (the time interval from the time that 5\% of the fluence
is accumulated to 95\%) range between 10 and 200 sec
(see Heise et al. 2000, Fig.~3).

\subsubsection{Search for FXTs as afterglows of GRBs} 

After the identification (Costa et al. 1999) of x-ray afterglows in GRBs, 
Grindlay (1999) suggested that a fraction of the FXTs might be such 
x-ray afterglows. 
The approximate agreement in rates, and 
derived log $N$- log $S$ between fast Transients and GRB afterglows would
rule out strong beaming differences (e.g., factors of $>3-10$) 
between prompt $\gamma$-rays of GRB and x-ray afterglows.

\subsubsection{Search for GRB afterglow in the ROSAT All Sky Survey.} 

Greiner et al. (2000) have searched for GRB x-ray afterglows in the ROSAT all-sky
survey in the energy range 0.1-2.4 keV. 
They find 22 afterglow candidates, where about 4 are predicted on the
basis of the same beaming angle in the x-ray and gamma-ray bands. The dwell
time on a source during the scan is typically 10-30 sec.
Follow-up spectroscopy strongly suggested a flare star origin in many,
if not all, cases.

\section{Fast Transient X-ray sources seen in the WFCs}
\subsection{Two types of Fast Transients}
Apart from the previously discussed transients, the X-ray transients observed
in the WFC are Fast X-ray Transients. The total operation period
of more than 5 years with the WFC have produced a homogeneous database 
with sufficient positional accuracy to search for identifications. 
A total of 39 sources have been detected up to Januari 2001.
They are seen at positions including high galactic latitude.
The observed sky distribution is consistent with being isotropic
(see Fig.~3).

There appears to be two classes of Fast X-ray Transients with different
outburst timescale.
A histogram of durations shows that the WFC-Fast Transient have a bi-modal
distribution with a peak around a few minutes (17 sources)
and a peak around an hour (22 sources).
A subset of the short ones are X-ray counterparts of GRBs. 
We called the remaining short duration sources
X-ray flashes (Heise et al. 2000).

The long duration FXTs typically between $2\times 10^3$ and $2\times 10^5$ s.  
In the latter class 9 out of 22 sources have been identified with galactic coronal
sources (6 flare stars and 3 RS-CVn variables). We do not know the
identity of the remaining sources on times scales of an hour, but 
we consider it probably that all are coronal sources. Such time scales
are indeed observed in dedicated studies of coronal sources.
In the remaining part we concentrate on the x-ray flashes.

\subsection{Properties of X-ray flashes}
\subsubsection{Definition, light curves and spectra.}

An operational definition of an X-ray flash in the WFC is a
FXT with duration less than a few thousand seconds,
which do not trigger $\gamma$-ray bursts experiments.
They are not detected by the Gamma Ray Burst Monitor
(GRBM) in the range 40-700 keV. 
This definition excludes the x-ray counterparts of the typical 
Gamma Ray Bursts as observed with BeppoSAX, 
which we will refer to as classical GRBs. This holds for
X-ray rich GRBs as well.

17 x-ray flashes have been observed in the WFC in about
5 years of BeppoSAX operations. 
They occur at a rate of one third of 
the x-ray counterparts of GRBs in the same instrument. 
Flashes are clearly distinct from GRBs in a plot of x-ray flux or fluence
versus $\gamma$-ray flux or fluence (Fig.~6). 
Gamma ray Bursts that are X-ray rich, such as GRB980326 and
GRB981226, have a $\gamma$-ray flux on the low side of the
distribution in Fig.~6, but still above the x-ray flashes.
They may be considered as bridging cases between GRBs and x-ray flashes.

X-ray flashes thus are bright x-ray sources with peak fluxes in the range
$10^{-8}$ and $10^{-7}$ erg/s/cm$^2$, 
the same range as the x-ray counterparts of classical GRBs.
An example of the light curve of an x-ray flash in different
energy bands is given in Figure~5. 
The energy spectra in the range 2-25 keV fit with a single power law 
photon spectrum and absorption consistent with galactic absorption.

The spectra are power law like (which distinguish the flashes from type~I
x-ray bursts). The power law photon index ranges between very soft spectra 
with photon index 3 to hard spectra with index 1.2.
We extrapolated such power law spectra into in the 40-700 keV range.
For at least two flashes, the GRBM upper limit is not consistent
with the extrapolation of the spectrum with the same powerlaw index,
indicating that a spectral break must occur in the energy range between 
the WFC and GRBM, typically between 30 to 50 keV.

\subsubsection{Low energy BATSE detections.}
The BATSE energy range extends to lower energies than the GRBM. 
10 out of 17 were potentially observable with BATSE and 9 out of these 10 
are actually detected in either the lowest or the lowest two BATSE energy channels, 
resp.\ 25-50 keV and 50-100 keV (Kippen et al. 2000). 
These events did not trigger the BATSE instrument,
but the sources are seen in the standard accumulations in 1 s 
timebins.
The 50-300 keV peak flux is near the threshold of detectability in BATSE,
whereas the 2-10 keV flux is bright.
The 50-300 keV fluences range between $5\times10^{-8}$ and $4\times10^{-7}$ 

The ratio of the 30-500 keV BATSE peak flux is displayed against the 2-10 keV
WFC peak flux in the left panel of Figure~6 and the same
ratio of the fluences in the right panel. 

\subsubsection{X-ray to $\gamma$-ray flux and fluence ratios.}
A number of classical GRBs observed in the WFCs are also detected by BATSE.
The ratio of peak flux and fluence is shown in Fig.~3, Heise et al. 2000,
for the two classes: 17 BATSE-detected bursts in the WFC and 
9 BATSE-detected x-ray flashes.
These ratios for the x-ray flashes extend the range
observed in normal GRBs by a large factor. 
Peak flux ratios of x-ray flashes extend up to a factor of 100,
and fluence ratios extend up to a factor of 20.

\subsection{Origin of X-ray flashes}
X-ray flashes could be GRBs at large cosmological redshift $z>5$, 
when gamma rays would be shifted into the x-ray range and the
typical spectral break energy at 100 keV shows at 16 keV.
A typical average spectrum of a GRB (spectral slope $\alpha=-1$
before the break, a break energy of 100 keV and a spectral slope
$\beta=-2.25$ after the break) would have a ratio of x-ray over
$\gamma$-ray fluence of around a few percent. This ratio becomes
larger than one at large redshift: 
a typical $\gamma$-burst becomes an x-ray flash. 
Note, however, that the spread in spectral properties of GRBs is large.
Extrapolations of these spectra into the x-ray range would predict
a large variation in the ratio of x-ray to $\gamma$-ray fluence.
These intrinsic variations are much larger than the average
increase with redshift.

A time dilation in the duration of x-ray flashes is not observed, 
nor has that been seen within the class of classical GRBs.
The intrinsic distribution of durations might dominate the effect caused by
cosmological redshift. Also observational selection effects in the
determination of the duration may be large. 
Moreover, if the average decay in time of the prompt emission 
has a power law shape (such as is the case for the afterglow
emission which in some cases smoothly joins the later stages
of the prompt emission), no obervation time dilation is seen in the
durations.

The statistical properties of X-ray flashes display many of  
the properties of GRBs, except that they are not readily seen at
$\gamma$-rays and do not trigger GRB-instruments. 
X-ray flashes therefore may show an extension of the
physical circumstances which lead to relativistic expansion
and the formation of GRBs, the process called the
cosmic fireball scenario.
Interestingly, Katz (1994), while discussing the cosmic fireball
scenario with extreme relativistic motion (large bulk Lorentz factors $\Gamma$), 
remarks that one should look for x-ray bursts and uv-bursts, because
such bursts are expected at mildly relativistic expansion at
lower bulk Lorents factor.
$\Gamma$ depends on the fraction of restmass
energy (the baryon load) to the total energy of the burst. 
A low baryon load leads to high Lorentz factors and probably high
energy GRBs.
A high baryon load (also called a "dirty fireball") leads to  a
smaller Lorentz factor and presumably a softer GRB:
possibly the x-ray flash.

In almost all progenitor models for GRBs, the GRB is produced 
by the final collapse of an accretion torus around a recently formed
collapsed object (black hole). In many cases the stellar debris around
the birthgrounds of GRBs prevents one from observing any
prompt high energy radiation at all, making GRBs a rare phenomenon as compared
to supernovae. It seems plausible that x-ray flashes bridge
the stellar collapses in a relatively clean direct circumburst environment
observed as normal GRBs and stellar collapses unobserved in x- and $\gamma$
radiation.

\end{document}